 \documentclass[final,5p,times,twocolumn,sort&compress]{elsarticle}

\usepackage{natbib}
\usepackage{amssymb}
\usepackage{lipsum}
\usepackage{epstopdf}
\usepackage{amsthm,amsmath}
\usepackage{hyperref}
\usepackage{xurl}
\usepackage{graphicx}
\usepackage{xcolor}
\usepackage{multirow}
\usepackage[hang,flushmargin]{footmisc}
\usepackage{lineno}
\setcitestyle{super,open={},close={}}

\makeatletter
\def\Hy@Warning#1{}
\def\blfootnote{\xdef\@thefnmark{}\@footnotetext}
\makeatother

\journal{Environmental Health Perspectives}

\begin{document}

\begin{frontmatter}



\title{Creating a Spatial Vulnerability Index for Environmental Health}


\author[first]{Aiden Price}
\author[second]{Michael Rigby}
\author[third]{Paula Fi\'evez}
\author[first]{Kerrie Mengersen}
\affiliation[first]{organization={Centre for Data Science, Faculty of Science},
            addressline={Queensland University of Technology}, 
            city={Brisbane City, Brisbane},
            postcode={4000}, 
            state={QLD},
            country={Australia}}
\affiliation[second]{organization={Australian Urban Research Infrastructure Network},
            addressline={University of Melbourne}, 
            city={Parkville, Melbourne},
            postcode={3052}, 
            state={VIC},
            country={Australia}}
\affiliation[third]{organization={FrontierSI},
            city={Docklands, Melbourne},
            postcode={3008}, 
            state={VIC},
            country={Australia}}

\begin{abstract}
Extreme natural hazards are increasing in frequency and intensity. These natural changes in our environment, combined with man-made pollution, have substantial economic, social and health impacts globally. The impact of the environment on human health (environmental health) is becoming well understood in international research literature. However, there are significant barriers to understanding key characteristics of this impact, related to substantial data volumes, data access rights and the time required to compile and compare data over regions and time. This study aims to reduce these barriers in Australia by creating an open data repository of national environmental health data and presenting methodology for the production of health outcome weighted population vulnerability indices related to extreme heat, extreme cold and air pollution at various temporal and geographical resolutions.

Current state of the art methods for the calculation of vulnerability indices include equal weight percentile ranking and the use of principal component analysis (PCA). The weighted vulnerability index methodology proposed in this study offers an advantage over others in the literature by considering health outcomes in the calculation process. The resulting vulnerability percentiles more clearly align population sensitivity and adaptive capacity with health risks. The temporal and spatial resolutions of the indices enable national monitoring on a scale never before seen across Australia. Additionally, we show that a weekly temporal resolution can be used to identify spikes in vulnerability due to changes in relative national environmental exposure.
\end{abstract}

\end{frontmatter}

\blfootnote{Address correspondence to Aiden Price, Centre for Data Science, Faculty of Science, Queensland University of Technology, Brisbane City, Brisbane, 4000, Queensland, Australia. Email: a11.price@qut.edu.au.}
\blfootnote{The authors declare they have no conflicts of interest to disclose.}
\blfootnote{No individual subject data have been acquired for this study. The authors obtained human ethics exemption for this study (ref. HE-Ex2 2022-4825-7319).}
\blfootnote{Conclusions and opinions are those of the individual authors and do not necessarily reflect the policies or views of EHP Publishing or the National Institute of Environmental Health Sciences.}




\section{Introduction}
\label{introduction}

In 2016, the World Health Organisation (WHO) stated that 13.7 million deaths a year, or 24\% of all global deaths, are linked to adverse environmental factors such as poor air quality, unstable climate and the built environment \cite{pruss2016preventing}. It is well understood that poor air quality and climate extremes have significant adverse impacts on human health \cite{seltenrich2015a,manisalidis2020a,i2021a,kinney2008a}. Even in Australia, where air quality is generally among the cleanest in the world \cite{iqair2021a}, pockets of poor air quality are said to be responsible for the deaths of approximately 3000 people per year \cite{unknown2011b}. Extreme heat is the focus of a large body of research investigating the relationship between climate and human health, with heatwave events contributing to excess mortality in Australia more than all other natural hazards combined since 1844 \cite{coates2014exploring,coates2022heatwave}. Cold-related mortality is also significant both internationally \cite{berko2014a, gasparrini2015a} and in Australia \cite{vardoulakis2014a}, especially when considered alongside populations with increased respiratory conditions \cite{wondmagegn2021a, kinney2015a}. These exemplar environmental exposures and their relationship with human health form the study of environmental health \cite{unknown-a}.

Environmental health is also affected by the ways in which cities are designed and built, as these factors can lead to populations being exposed to or protected from extremes in the natural environment \cite{frank2005a}. Moreover, population characteristics may determine areas which are more vulnerable to natural hazards than others, both socially and economically \cite{statistics2016a}. As the frequency and intensity of extreme climate events have increased in recent years, greater emphasis is being placed on adaptation and mitigation strategies which consider environmental factors as well as population socio-demographics and the impact of urban infrastructure, both in Australia and globally \cite{nswsote2021,d2010a,spickett2008a,armstrong2018human,unknown-g}.

Historically, significant efforts have been made to characterise the human and financial risks related to natural hazards. More recently, these efforts and the corresponding risks have become focused more sharply on the vulnerability of exposed human populations rather than the exposure itself \cite{burton2018a}. The term vulnerability, in social science research, generally describes a state of people and populations which varies significantly both geographically and historically \cite{wisner1994a}. The changing nature of many hazards, coupled with growing and ageing populations and infrastructure in exposed areas is leading to increased vulnerability across Australia and internationally \cite{taskforce2018a}.

Due to the complex relationships between the environment and humans, no single vulnerability measure will provide a holistic and comprehensive indicator of the state of environmental health vulnerability within a population, cohort, or community of interest \cite{dwyer2004a}. Despite this, local government agencies need information to support vulnerability assessments in order to effectively develop policy, apply intervention, provide emergency disaster management and improve their respective community’s ability to adapt to changes in environment. Effective adaptation and risk management strategies and practices require an in-depth understanding of population vulnerability, as well as the various exposures facing populations. In addition, it is important to provide an evidence-based assessment of changes in these areas \cite{climate2012a}, as ``hazards only lead to disaster if they intersect with an exposed and vulnerable society and when the consequences exceed its capacity to cope" \cite{taskforce2018a}.

It is especially difficult to assess environmental health risks in Australia, due to enormous variations in geography, climate and environmental burden of disease \cite{begg2007burden, beaty2021a}, as well as the continent's extremely heterogeneous population density and socio-demographics. For instance, Australian regions are located within large latitude and longitude ranges, between $10^\circ$S to $45^\circ$S and $113^\circ$E to $153^\circ$E, respectively. In addition, this landscape consists of largely remote regions, with over 68$\%$ of Australians living in the greater metropolitan area of Australia's eight capital cities in 2021 \cite{AIHWpop2022}. Decision making at the local area level is split between the federated national, state and local government political system, with policy often informed by national averages, instead of relevant local characteristics, e.g., area level health, socio-economics and location characteristics \cite{parry2007climate}.

The analysis of health outcome data is approached in two main ways in environmental health literature. The first approach is a typical statistical analysis, in which environmental variables are assessed against one or more health outcomes of interest \cite{b2010a,keijzer2017a,hondula2012a}. The second approach is the development of environmental health indicators or indices, which are based on relevant literature, rather than being directly assessed against specific health outcomes \cite{rinner2010a,reid2009a,yu2021a,el-zein2015a,new-a,inostroza2016a}. Environmental health indicators are parameters that provide information about the state of the environment and its link to human health \cite{unknown-a}. A composite indicator or index, on the other hand, is a measure that combines multiple parameters or indicators in order to provide an overall assessment of the environmental health status of a particular area or population, often summarising underlying information via a single value or ranking \cite{brousmiche2020spatialized}. Transforming or combining raw data into indicators and indices enables vulnerable regions to be more easily identified so that their underlying characteristics may be further assessed and options for policy intervention and natural hazard response can be more readily explored. As a result, indicators and indices have become popular among state governments and environmental health organisations to communicate environmental health risk in Australia \cite{nswsote2021,d2010a,spickett2008a}.

This paper utilises a range of environment and health data to provide an assessment of environmental health risk through the production of vulnerability indices for two case studies across Australia: extreme climates (heat and cold) and air quality. These environmental health domains were prioritised by a large group of relevant government organisations across a range of domains such as water quality and soil contamination. The vulnerability indices produced in this work cover various spatial and temporal resolutions, thereby capturing seasonal and long-term changes in population vulnerability.

The vulnerability indices produced in this paper have been implemented into the Australian Environmental Health (AusEnHealth) decision support platform. AusEnHealth enables the monitoring and analysis of over 100 variables from the many sectors which contribute to environmental health risk, providing insight into the impacts of the changing environment on human health in Australia. The spatial resolutions captured in AusEnHealth include statistical areas (SA2, SA3 and SA4) and local government areas (LGA). Statistical areas are a hierarchical construct developed to reflect the location of people and communities and are updated every five years as part of the Australian Statistical Geography Standard (ASGS) based on the Australian Census \cite{abs2021australian}. Local government areas on the other hand, represent boundaries that relate to the administration of cities, shires, municipalities or towns \cite{LGAabs2023, LGAcsiro2023} within Australian states and territories

This paper is set out as follows. Section \ref{s:Methods} introduces the data considered for use in the creation of vulnerability indices and covers both the existing and proposed methodology used to create vulnerability indices. This section also explains data considered by similar works in the literature, as well as spatio-temporal resolutions and extents. Section \ref{s:Results} presents the main result of this study and outlines the differences between the existing methodology and the proposed weighted methodology. In addition, this results section reiterates the value of the overall data assets curated and created for this study by demonstrating how vulnerability indices can be pulled apart to more carefully explore and understand exposed or sensitive populations. In Section \ref{s:Discussion}, the benefits of the proposed methodology are discussed from a health risk perspective, with particular attention paid to both the granularity of the indices as well as the ability to observe trends over time. Finally, section \ref{s:Conclusion} summarises the study and suggests potential future direction.

\section{Methods}\label{s:Methods}
\subsection{Data}\label{s:Data}
The data used in this study have been collected as part of the AusEnHealth initiative, which involved a national data audit of Australian environmental health data at both state and national levels. During this process, data custodians, datasets and their related metadata were compiled and the available data were assessed for national coverage, relevance, spatial resolution and accessibility.

This study features data related to extreme climates and air pollution which were the environmental health domains prioritised by collaborating government organisations. Ethics for data collection and storage were approved as part of a human ethics exempt research application at the Queensland University of Technology (reference HE-Ex2 2022-4825-7319).

The remainder of this section provides an overview of the variables and data custodians considered to effectively capture risk related to the priority environmental health domains. Following methods related to the data, we provide an overview of existing vulnerability index methodology as well as the modifications made to produce health outcome weighted vulnerability indices.

\subsubsection{Climate}
Climate data relevant to the extreme climates case study comprised minimum and maximum temperature (Table \ref{D_Clim}), sourced from the Bureau of Meteorology (BOM) over a 20-year period \cite{meteorology2021a}. These climate data were used to create two indicators related to the impact of temperature on human health, namely excess heat factor (EHF) and excess cold factor (ECF) using existing methodology \cite{nairn2013a,nairn2015a}. EHF and ECF are metrics used by BOM and commonly in literature \cite{beaty2021a,scalley2015a,scalley2014a,xiao2017a} to determine whether an area is impacted by a heatwave or coldwave, respectively, using short- and long-term historical temperature recordings.

\subsubsection{Air Quality}
Relevant air pollutant data were collected from the Copernicus Atmosphere Monitoring Service (CAMS) \cite{copernicus2021a}. Air quality was determined by the measurement of air pollutants which are introduced into the air via road traffic, industrial processes and bushfires. The resultant variables comprised five chemical compounds and three sizes of particulate matter (Table \ref{D_Air}).

\subsubsection{Built Environment}
Primary built environment data used in this study are provided in Table \ref{D_BE} and were acquired from the Australian Bureau of Statistics (ABS) \cite{statistics2021a}, Geoscience Australia's Digital Earth Australia (DEA) \cite{australia2021a}, NASA's Moderate Resolution Imaging Spectroradiometer (MODIS) \cite{n2021a} and the Terrestrial Ecosystem Research Network (TERN) \cite{t2021a}. The variables accessed include the number of hospitals, green space percentages, water surface percentages, the normalised difference vegetation index (NDVI) and canopy cover percentages.

\subsubsection{Demographics}
Demographics data have been utilised in numerous studies to determine vulnerable populations \cite{flanagan2011a,cutter2003a,statistics2001a}. Twenty relevant demographic variables were obtained from the Australian Census data \cite{statistics2021a}. In addition, eleven variables related to existing health conditions were sourced from the Public Health Information Development Unit (PHIDU) social health atlas \cite{unknown2020b, unknown2021a}. These data are displayed in Table \ref{D_D}.

\subsubsection{Mortality}
This study utilises publicly available mortality data sourced from the mortality over regions and time (MORT) books compiled by the Australian Institute of Health and Welfare (AIHW). This data set contains numbers of deaths by leading causes of death disaggregated by sex across Australia \cite{unknown2021b}. Relevant health outcome variables from this data set are given in Table \ref{D_HO}. Leading causes of death have been grouped to to produce mortality data related to heat, air quality, or all causes, determined by combining or referring to specific mortality causes listed in Table \ref{T:ACM}.

\subsubsection{Data Sourcing and Processing}
The data were most commonly downloaded manually from the sources using file transfer protocol (FTP). In some cases, datasets were only available in a series of files, e.g., yearly, monthly, or daily datasets. Automated downloads were required for these and relied on Hypertext Transfer Protocol (HTTP) requests to online data storage servers using the statistical software R \cite{R2020}. All data used in the creation of the vulnerability indices classify as geographic or spatial data, i.e., associated with a location relative to Earth. Where only gridded or raster data were available, these were transformed using R or GIS into spatially aggregated datasets with consistent spatial resolutions. Some data sets are stored as XLSX or CSV files with additional metadata items included in the data entries, where available. In these instances, variables were extracted and matched with a standard set of geographical resolutions codes to ensure consistency.

All data types were accessed at their finest temporal resolution and have been aggregated where needed to produce yearly, monthly and weekly datasets. The spatial resolutions utilised in this study include ASGS statistical areas level 2-4 (SA2, SA3, SA4), as well as local government areas (LGA). The number of SA2, SA3 and SA4 regions in Australia are 2310, 358 and 107, respectively, with SA2 representing the highest resolution. There are also 566 LGAs in Australia, with this resolution being included for its relevance to policy planning and the decision support component of this study and the overall AusEnHealth initiative. Both the processed data and the associated metadata were stored in Amazon S3 buckets, maintained on servers at the Queensland University of Technology.

While linked individual data could add significant value in areas of risk management, disaster preparedness and policy development, such data could not be accessed for this work, due to legal, ethical and governance barriers in accessing socio-demographic and health outcome data from the Australian Bureau of Statistics and the Australian Institute of Health and Welfare. The release of individual data must also comply with privacy laws and regulations specific to each jurisdiction (Australian State or Territory). Hence, aggregated data were instead used to enable the creation of vulnerability indices at the national scale. This aligns with many other national and international atlases \cite{unknown-e, australia-a, ASPE2021, SAHSU2021, CDC2022, EPA2023}.

\subsection{Vulnerability Indices}\label{vuln}
The national vulnerability indices created in this study utilise a consistent environmental health vulnerability index framework introduced by the Intergovernmental Panel on Climate Change (IPCC) in 2007 \cite{parry2007climate}. This framework aims to better explain population vulnerability with respect to climate and consists of three components: exposure, sensitivity and adaptive capacity. Exposure is a direct measure of household, community, or population exposure to a certain event (extreme heat, cold, air pollution, etc.). Sensitivity is defined as the susceptibility of a household, community, or population to the exposure. Adaptive capacity captures the capabilities of a household, community, or population to cope with or recover from the impact of the exposure. In the creation of a vulnerability index, these components form sub-indices which contain relevant variables and indicators. The sub-indices are used to capture different aspects of vulnerability and are ultimately combined into an overall vulnerability index: a single measure of population vulnerability with respect to a particular exposure event.

\subsubsection{Vulnerability Index Components and Extent}
A number of vulnerability indices exist which focus primarily on population variables. Prominent $21^{st}$ century examples of these indices include the social vulnerability index (SVI) \cite{flanagan2011a}, the social vulnerability index for the United States (SoVI) \cite{cutter2003a} and Australia's index of relative socio-economic disadvantage (IRSD) \cite{statistics2001a}. While the composition of each vulnerability index differs, common variables include employment status, high school education, English proficiency and income. These social vulnerability indices, as well as resilience and risk indices, have since become widely known for providing insight into disaster response, recovery, readiness, adaptability, resilience and risk assessment \cite{statistics2001a}.

For AusEnHealth, the population vulnerability stemming from demographic variables was combined with vulnerability associated with exposure to certain environmental hazards. The variables considered feasible for use in the creation of AusEnHealth heat, cold and air quality vulnerability indices rely on existing literature \cite{beaty2021a,yu2021a,el-zein2015a,new-a,inostroza2016a} which utilise the accepted exposure, vulnerability and adaptive capacity framework \cite{parry2007climate}. While vulnerability index development in recent literature has a heat-health focus, similar data can be used to produce a cold vulnerability index. There is extensive literature \cite{ghose2004a,romero-lankao2013a,dasgupta2021a} from which to draw information for the development of an air quality vulnerability index. The components used to create the AusEnHealth vulnerability indices are listed in Tables \ref{T:E}, \ref{T:S} and \ref{T:A} and have been separated by exposure, sensitivity and adaptive capacity components, respectively.

Of particular note, Table \ref{T:E} introduces historical temperature percentiles, which were calculated by comparing each region to its own historical values. This relative calculation of temperature allows for the resulting indices to capture spikes in heat vulnerability in regions which are typically cooler or hotter by accounting for acclimatisation \cite{nairn2013a}.

Interestingly, no two vulnerability index implementations in the literature use the same set of variables. Reasons for these differences include regional scope, data availability, application and differences in the groupings of exposure, sensitivity and adaptive capacity. This also applies to the AusEnHealth vulnerability indices. Heatwave and coldwave measures were included in the climate exposure sub-index to incorporate additional information relating to consecutive high heat or cold days into the overall vulnerability indices. As noted above, the temperature measures here are given as historical percentiles to represent out-of-the-ordinary heat on a per-region basis. As there is no identified acclimatisation to poor air quality in the literature, only spatial percentiles were considered for the air quality index. The air quality sub-index only includes air pollutants, following the same approach as the heat and cold exposure sub-indices. Sensitivity variables were largely based on those used in published population vulnerability indices \cite{flanagan2011a,cutter2003a,statistics2001a}, with particular influence from the recent heatwave report from BOM \cite{beaty2021a}. Selection of variables for adaptive capacity centred around population mobility and access to services, such as hospital counts, access to internet and the number of dwellings with at least one vehicle.

The majority of vulnerability indices identified in the research literature combine data from a wide temporal range to form a single index \cite{beaty2021a,yu2021a}. When changes over time are considered in the creation of vulnerability indices, this is usually over either a short time period or only a few time points \cite{new-a}. The AusEnHealth vulnerability indices are presented at weekly, monthly and yearly scales by incorporating daily climate and air quality data and considering changes over time to the exposure, sensitivity and adaptive capacity sections of the created vulnerability indices. As the temporal extent of collected demographic data covers the years 2011-2019 (see Section \ref{s:Data} and Table \ref{D_D}), the vulnerability indices are limited to this time period.

As demographics data are not available at a weekly temporal resolution, data from a given year were used at each month and week interval in finer resolution data sets so that climate and air quality data could be included at those finer resolutions to show temporal variations in population vulnerability otherwise obfuscated by longer term averages. In consideration of variation over time, data were imputed using each region's temporal data where available. For example, unemployment rate is available only at yearly temporal resolution in 2011 and 2016, so it was imputed first annually for the years 2012 through 2015 and then replicated for each month and week entry in a given year for a given region. No spatial imputation was considered, due to large spatial heterogeneity. If missing data from a region could not be imputed using time-series data, that entry remained missing. The following section includes details of how this missing data were handled in the creation of vulnerability indices.

\subsubsection{Approaches to Vulnerability Index Methodology}
There are two prominent approaches to index creation. The first is a percentile rank-sum approach \cite{flanagan2011a}, which transforms variables into percentiles ranked from highest to lowest inferred vulnerability across all assessed regions. For a given region, the percentile values corresponding to the set of variables are then summed arithmetically with equal weights to produce the vulnerability index for that region.

The conceptual justification behind the use of equal weights in the above approach is that when trying to cover the broad spectrum of exposure-related health outcomes, weighting variables by comparing against any single health outcome is likely to bias the vulnerability index compared to a separate exposure-related health outcome. However, the disadvantage of the equal weights approach is that some of the variables may be more important across a wide range of health outcomes and care must be taken to ensure that all variables correlate with population vulnerability. Moreover, if health outcomes are not factored into the creation of the index and equal weights are applied, the resulting vulnerability index may not be adequately correlated with any or all health outcomes. This means that a vulnerability index created in this way in different places in the world or with different variables will not reliably capture the same population vulnerabilities. Additionally, substantive dependencies between variables are not taken into account, which may result in highly correlated variables controlling the overall relationship of the vulnerability index.

The second approach weights and combines variables using a dimension reduction approach such as principal component analysis (PCA) or its derivatives such as factor analysis \cite{cutter2003a,statistics2001a,conlon2020mapping}, where the first principal component (or factor) is taken as the vulnerability index. There are strengths and weaknesses of the PCA methodology. One strength is that PCA identifies patterns and associations between the different dimensions of the underlying data and can adjust for the contribution of multiple correlated variables. On the other hand, PCA does not compare the variables to a health outcome and so will produce weights that optimally describes variance in a set of variables that may not be correlated to any exposure-related health outcome. That is, depending on the quality (and quantity, perhaps more importantly) of underlying variables, PCA-derived weights may reduce the overall value of a vulnerability index compared to one created using equal weights. Moreover, a single principal component (or factor) may explain a relatively small amount of the total variation in the input variables.

Ultimately, the approach used to derive a vulnerability index, as well as the underlying components, must be based on data available within the underlying region(s). Recent discussions led by NASA \cite{nasa2022} conclude that chosen weights, correlated parameters and categorisation of exposure, sensitivity and adaptive capacity parameters are largely subjective and will depend on what is being monitored or evaluated.

A notable characteristic of both of the existing approaches is that neither are expressly linked to health outcome(s) of interest, other than by the initial choice of the variables included in the index construction. A strong link is often difficult to accomplish due to the complex associations and potential temporal lags between the underlying environment and health indicators, the complex spatial and temporal variation within these indicators and the overarching challenge of capturing general vulnerabilities, i.e., avoiding the creation of an index that is too prescriptive with respect to a single health outcome. 

This paper presents an alternative to the sum-rank and PCA approaches by weighting components using their pairwise correlation with a desired health outcome. The benefit of this approach is that the resulting vulnerability index is intrinsically but lightly associated at creation with the desired health outcome.

The following subsections detail both the equal weight and health outcome weighted methods.

\subsubsection{Equal Weight Vulnerability Indices}
The equal weights method \cite{flanagan2011a} is the baseline approach in the creation of vulnerability indices. In this approach, a regional sub-index is created by summing together spatial percentiles of its variables. Let $S_{ikt}$ be a sub-index of region $i$ at time $t$ containing $N_k$ variables, where $k=1,2,3$ corresponds to exposure, sensitivity and adaptive capacity sub-indices, respectively. With variables $n=1,\dots,N_k$, a vulnerability sub-index, $S_{ikt}$, for a geographical region, $i$, at time, $t$, is calculated as follows,

\begin{equation}
S_{ikt} = \sum_{n=1}^{N_k}f_i(\underline{x}_{nt}),
\end{equation}

\noindent where $\underline{x}_{nt}$ is a vector containing every region's observed value of variable $n$ at time $t$ and the function $f_i$ computes the $i^{th}$ region’s spatial percentile from $\underline{x}_{nt}$. The final vulnerability index, $VI_{it}$, is calculated in a similar way.

\begin{equation}\label{ew}
VI_{it} = \sum_{k=1}^{3}f_i(\underline{S}_{kt}),
\end{equation}

\noindent where $\underline{S}_{kt}$ is a vector containing every region’s observed value of vulnerability sub-index $k$ at time $t$.

The index detailed in Equation \ref{ew} will not produce an accurate result if, for a given variable n, the vector $\underline{x}_n$ contains missing values. Either the summation will produce a ``NaN" result, or the missing value(s) must be ignored, under-representing the value of the final index. A simplistic approach is to replace a missing value with the average of the existing data percentiles. To reduce the impact of highly correlated parameter spatial rankings and to ensure equal weighting of important aspects of vulnerability, this approach has also been modified to include sub-index themes, which act as sub-indices to the existing sub-indices \cite{beaty2021a}. Combined with the above calculations, this is equivalent to the following changes to the sub-index and vulnerability index equations.

\begin{equation}
S_{ikt} = \frac{1}{T_{k}}\sum_{p=1}^{T_{k}}f_i\Bigg(\frac{1}{N_{pt}}\sum_{n=1}^{N_{pt}}f_i(\underline{x}_{npt})\Bigg),
\end{equation}

\noindent where $N_{pt}$ is the number of variables in theme $p$ that contain at least one non-missing observation at time $t$, $\underline{x}_{npt}$ is a vector containing every region's value of variable $n$ within theme $p$ at time $t$ and $T_{k}$ is the number of themes in sub-index $k$. Note that $T_{k}$, like $N_{pt}$, is free to vary between sub-indices. Hence

\begin{equation}
VI_{it} = \frac{1}{3}\sum_{k=1}^{3}f_i(\underline{S}_{kt}).
\end{equation}

An alternative approach to handling missing values is to introduce some uncertainty in the imputation process. For example, uncertainty can be introduced through an adaptation of the familiar k-fold multiple imputation approach \cite{schafer1998multiple}. The average of the computed indices can then be taken if a single value of the index is required.

\subsubsection{Health Outcome Weighted Vulnerability Indices}
The second methodology employed in AusEnHealth utilises the pairwise correlation values between each variable and age-standardised all cause mortality rates to form a weighted average. Kendall’s Tau has been used over the typical Pearson correlation, as the linearity and normality assumptions for Pearson correlations may not be met for all variables. Note that weights can be positive or negative depending on the relationship between a particular variable and the health outcomes considered. The calculation of each weighted sub-index, $WS_{ikt}$, at time $t$ for region $i$ is as follows.

\begin{equation}
WS_{ikt} = \frac{1}{T_{k}}\sum_{n=1}^{T_{k}}f_i\Bigg(\frac{1}{N_{pt}}\sum_{n=1}^{N_{pt}}w_{npk}f_i(\underline{x}_{npt})\Bigg),
\end{equation}

\noindent where $w_{npk}$ is the Kendall’s Tau correlation between variable $n$ of theme $p$ and all-cause mortality in sub-index $k$. Note that equal weights are used for the exposure sub-index, i.e., $w_{np1}=1$ (see Discussion). The calculation of the overall weighted vulnerability index, $wVI_{it}$, remains the same, replacing the original sub-indices for the weighted sub-indices:

\begin{equation}
wVI_{it} = \frac{1}{3}\sum_{k=1}^{3}f_i(\underline{WS}_{kt}).
\end{equation}

\section{Results}\label{s:Results}
\subsection{Weighted Vulnerability Indices}
\begin{figure}[htbp!]
	\includegraphics[width=0.9\linewidth]{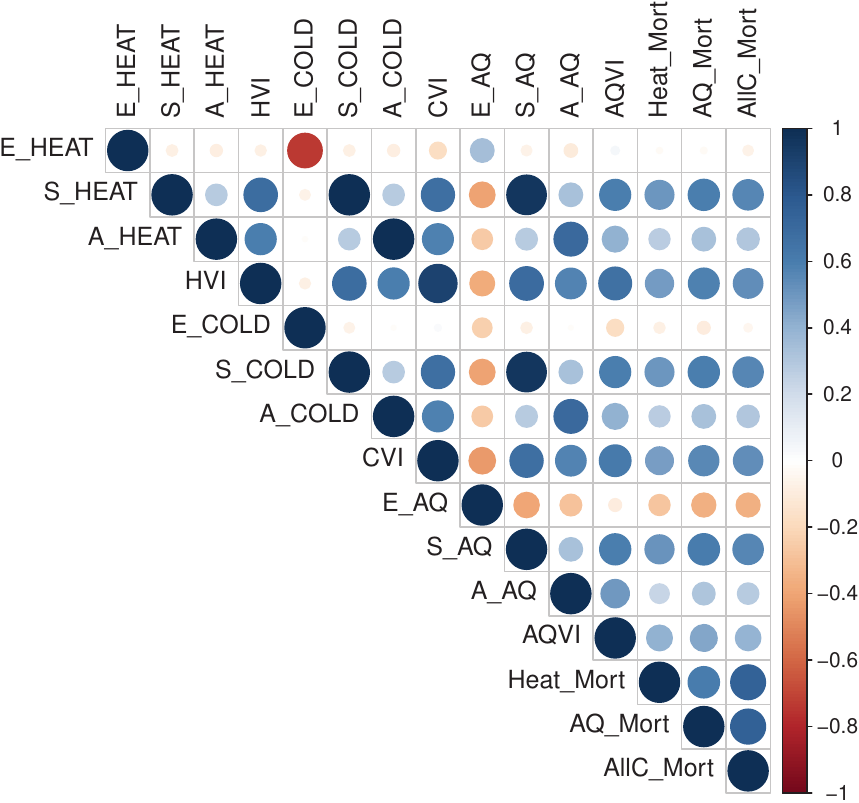}
	\caption{Pairwise correlation plot comparing health outcome weighted vulnerability index components against and health outcomes (heat, air quality and all-cause mortality). E, S and A correspond to exposure, sensitivity and adaptive capacity sub-indices, respectively.}
	\label{fig_vi_corr}
\end{figure}

\begin{figure}[htbp!]
	\includegraphics[width=0.9\linewidth]{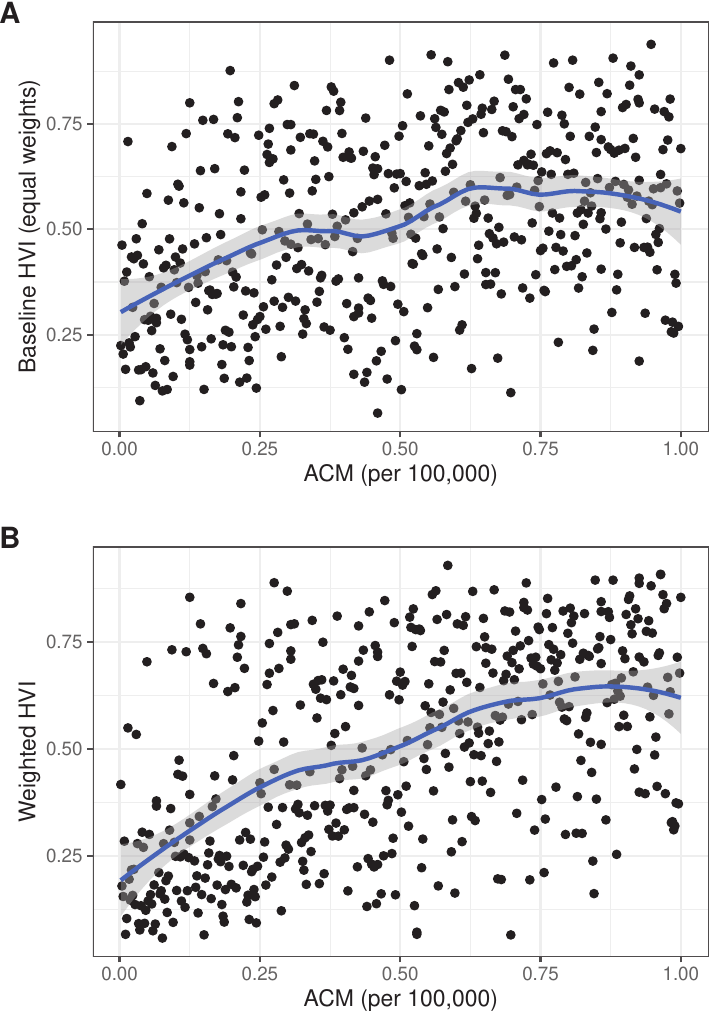}
	\caption{Scatter plot of age-standardised all cause mortality (ACM) against (A) the baseline heat vulnerability index (HVI) showing low correlation and (B) the weighted heat vulnerability index (wHVI) showing improved correlation (B).}
	\label{fig_corr_scatter}
\end{figure}

\begin{figure}[htbp!]
	\includegraphics[width=0.9\linewidth]{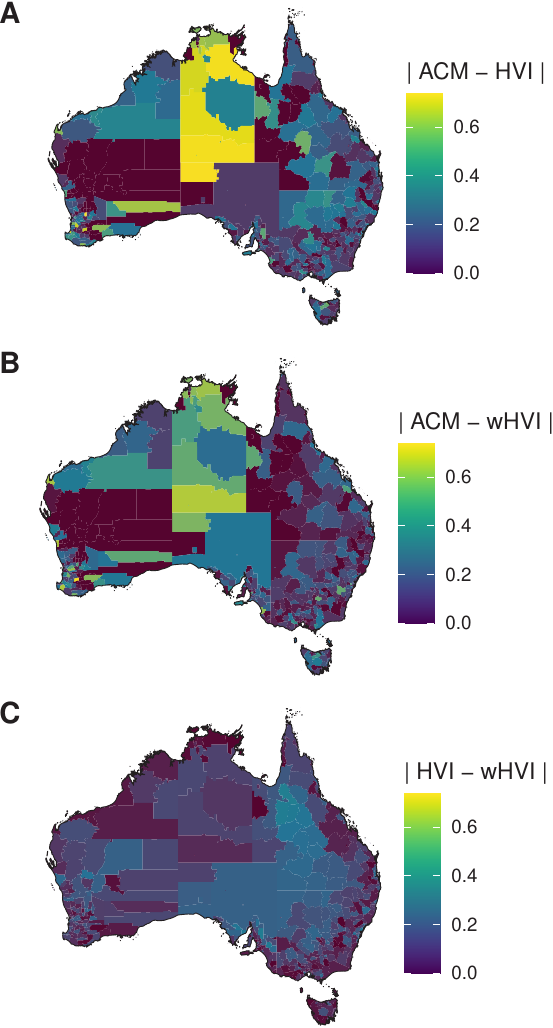}
	\caption{Choropleth map displaying the absolute difference between (A) age-standardised all cause mortality (ACM) and the baseline equally weighted heat vulnerability index (HVI), (B) ACM and the weighted heat vulnerability index (wHVI) and (C) HVI and wHVI.}
	\label{fig_corr_diff}
\end{figure}

The pairwise correlations (calculated using Kendall’s Tau) between the weighted vulnerability components and age-standardised all cause mortality data can be seen in Figure \ref{fig_vi_corr} for heat, cold, air quality, their sub-indices and three mortality categories (heat, air quality and all-cause mortality) over the 2015-2019 period. The improvement in correlation between the two vulnerability index methodologies and age-standardised all cause mortality is shown in Figure \ref{fig_corr_scatter}, where spatially ranked all cause mortality is plotted directly against both (A) the baseline heat vulnerability index, HVI and (B) the weighted heat vulnerability indices, wHVI. Choropleth maps additionally provide spatial insight into this improvement in correlation in Figure \ref{fig_corr_diff}, where the difference between spatially ranked all cause mortality (ACM) and heat vulnerability index methodologies (HVI and wHVI) are shown. The value of the weighted approach in Figure \ref{fig_corr_diff} is clear, showing a substantively smaller difference between ACM and wHVI than between ACM and HVI, both in urban and remote regions. Additionally, Figure \ref{fig_corr_diff} (C) shows the locations regions whose vulnerability changes the most between methodologies.

The value of presenting vulnerability indices at a higher temporal resolution is highlighted in Figure \ref{fig_vi_time} by displaying changes in heat vulnerability at the yearly, monthly and weekly temporal resolutions for the Derwent Valley LGA region in Tasmania. Derwent Valley is the 7$^{th}$ most vulnerable region in the 2015-2019 weighted heat vulnerability index and had the 58$^{th}$ highest all cause mortality over the same period. The weekly heat vulnerability index is shown over time and compared to the annual index in Figure \ref{fig_vi_2019} to reveal how seasons or events may impact a region's vulnerability.

\subsection{Comprehensive Data Asset}
Another outcome of this study is the creation of a significant data asset: almost a decade of time-series climate and air quality vulnerability indices at multiple geographical resolutions. To enable exploration of the data, an application has been created using the \texttt{shiny} package in R \cite{rshiny}, which offers flexible visualisations of both the equal and weighted vulnerability indices for heat, cold and air quality and their underlying components at their various spatial and geographical resolutions (see \url{https://qutcds.shinyapps.io/AusEnHealth_Interactive/}).

\subsection{Understanding Vulnerability}
\begin{figure*}[ht!]
	\includegraphics[width=0.95\linewidth]{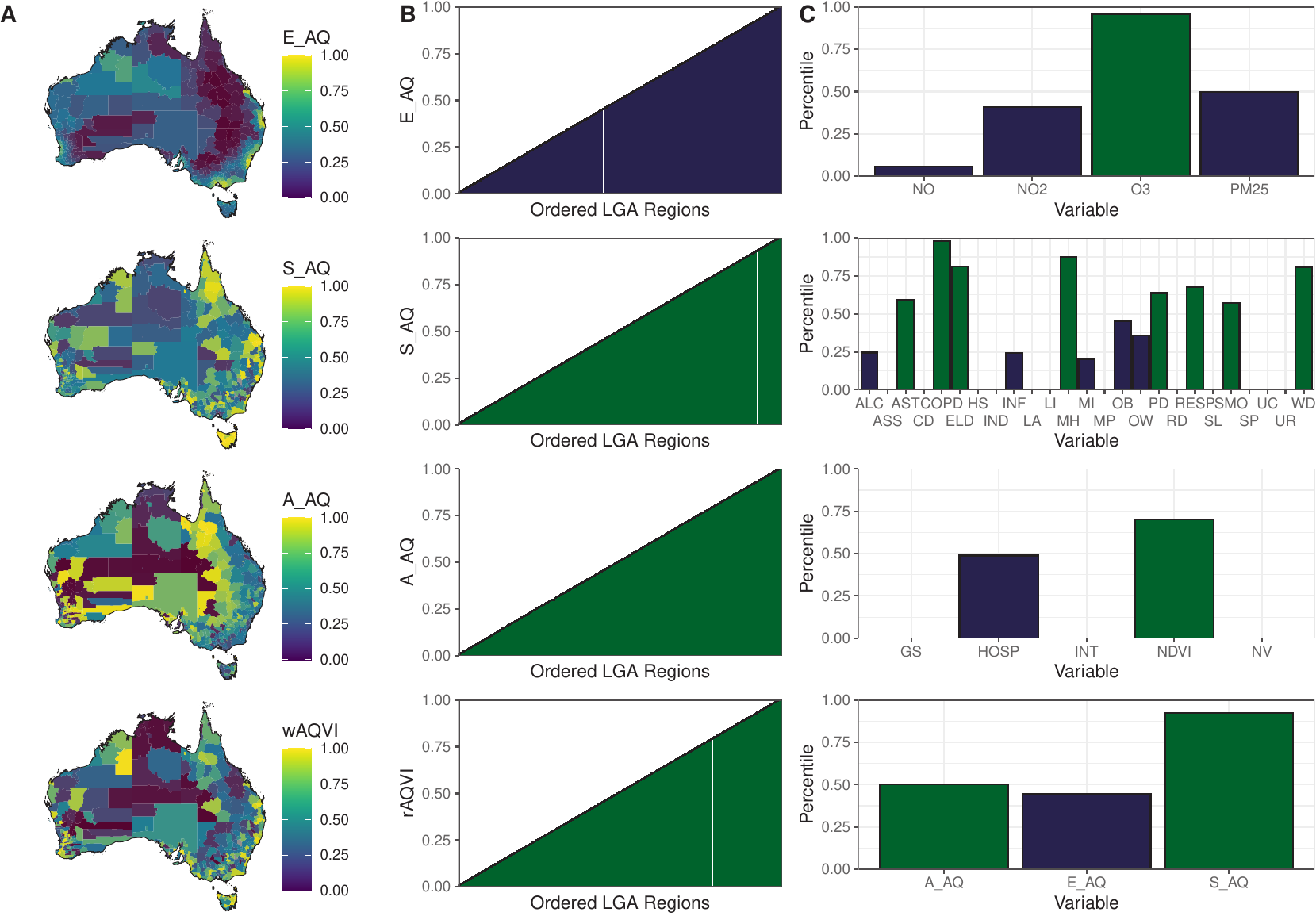}
	\caption{Breakdown of 2017 weighted air quality vulnerability index (wAQVI). Column A displays exposure, sensitivity, adaptive capacity and overall wAQVI indices. Column B displays ordered bar charts of the relevant index, with the Bundaberg LGA region marked white to indicate relative vulnerability among each index. Column C displays the spatial percentiles of all underlying parameters available for the Bundaberg LGA region in 2017, used to create each index. For Columns B and C, dark blue is used to represent values below the 50$^{th}$ percentile, with dark green representing values at or above the 50$^{th}$ percentile. Abbreviations used in this figure can be identified from the corresponding data tables \ref{T:E}, \ref{T:S} and \ref{T:A} for exposure, sensitivity and adaptive capacity, respectively.}
	\label{f:breakdown}
\end{figure*}

One disadvantage of merging variables and indicators in the creation of a vulnerability index is that it becomes unclear what makes a region vulnerable, especially when there are many underlying components. While the overall measure of vulnerability is valuable for relative assessments of regions, the provision of underlying data allows for particular pockets of vulnerable populations to be explored. To demonstrate, Figure \ref{f:breakdown} shows a breakdown of the weighted air quality vulnerability index, using the Bundaberg local government area (population density less than 15/km$^2$) to highlight the underlying vulnerability components in a regional city. It is clear from Figure \ref{f:breakdown} that the vulnerability of Bundaberg is heavily influenced by sensitivity, with high percentiles including percentage with chronic obstructive pulmonary disease (COPD), percentage with asthma (AST), percentage elderly (ELD), percentage living in mobile home (MH) and percentage with disability (WD). The underlying components in the Bundaberg LGA clarify why this particular population remains vulnerable with respect to air quality, despite a low exposure index. This breakdown process is just one way in which these indices can be used to better understand the factors which contribute to a vulnerable population.

\section{Discussion}\label{s:Discussion}
\subsection{New Approach to Vulnerability Index Creation}
Vulnerability indices are well known in the literature for their utility in emergency response \cite{flanagan2011a} and are emerging as a way to communicate the need for changes to policy, adaptation plans and intervention strategies by highlighting critical population vulnerabilities using a single value \cite{beaty2021a,nasa2022}. Despite the reputation of vulnerability indices in the literature, the strength of the relationship between most existing vulnerability indices and adverse health outcomes is not strong \cite{niu2021systematic}. The proposed inclusion of weights that are directly tied to targetted health outcomes (e.g., mortality data) improve the strength of this relationship and hence inherently increase the value of the indices for use in relevant risk assessment purposes in practice.

It is of particular importance to raise the temporal resolution of existing vulnerability indices for the purpose of communicating need. While it has been common historically and recently to combine multiple years of data in the creation of vulnerability indices \cite{beaty2021a,conlon2020mapping}, it is clear that fine temporal-scale extreme events are lost in such averages (see correlations between exposure sub-indices and mortality in Figure \ref{fig_vi_corr}). It is for this reason that no weight has been applied to the exposure sub-index in the vulnerability index methodology.

\begin{figure}[htbp!]
	\includegraphics[width=0.95\linewidth]{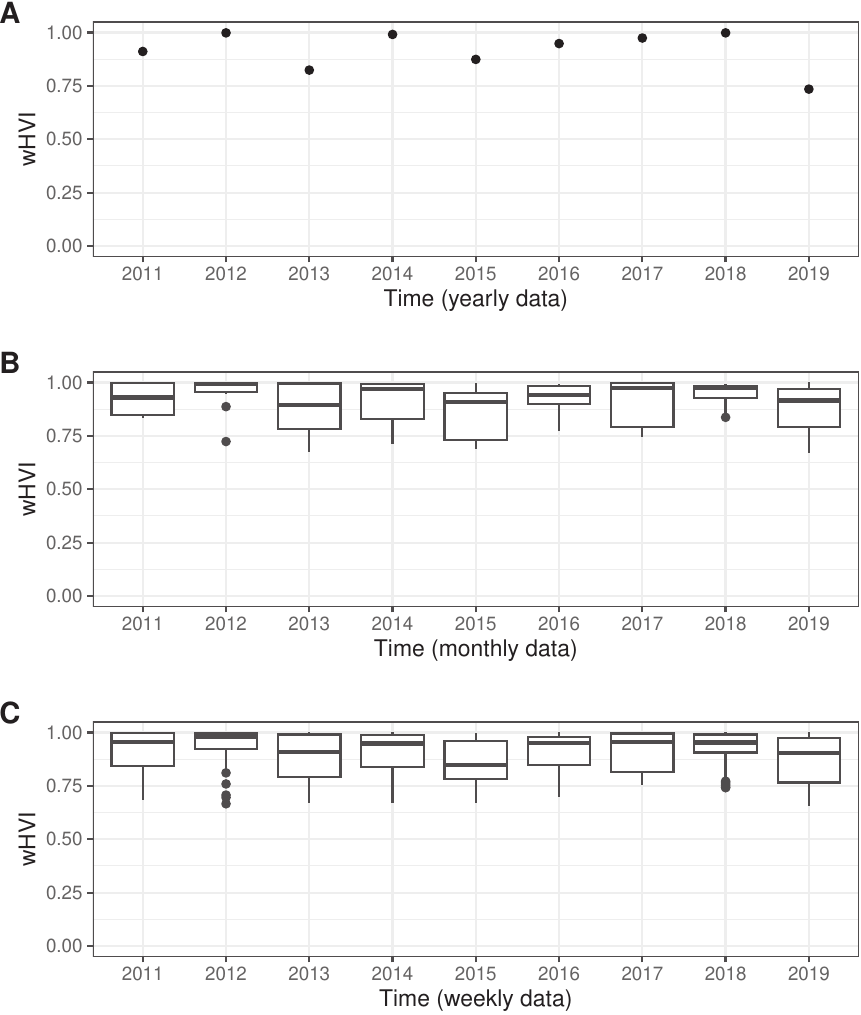}
	\caption{Representations of weighted heat vulnerability index (wHVI) calculations at different temporal resolutions. These are presented as (A) a time series plot of annual wHVI calculations, (B) yearly box plots of monthly wHVI calculations and (C) yearly box plots of weekly wHVI calculations.}
	\label{fig_vi_time}
\end{figure}

\begin{figure}[htbp!]
	\includegraphics[width=0.95\linewidth]{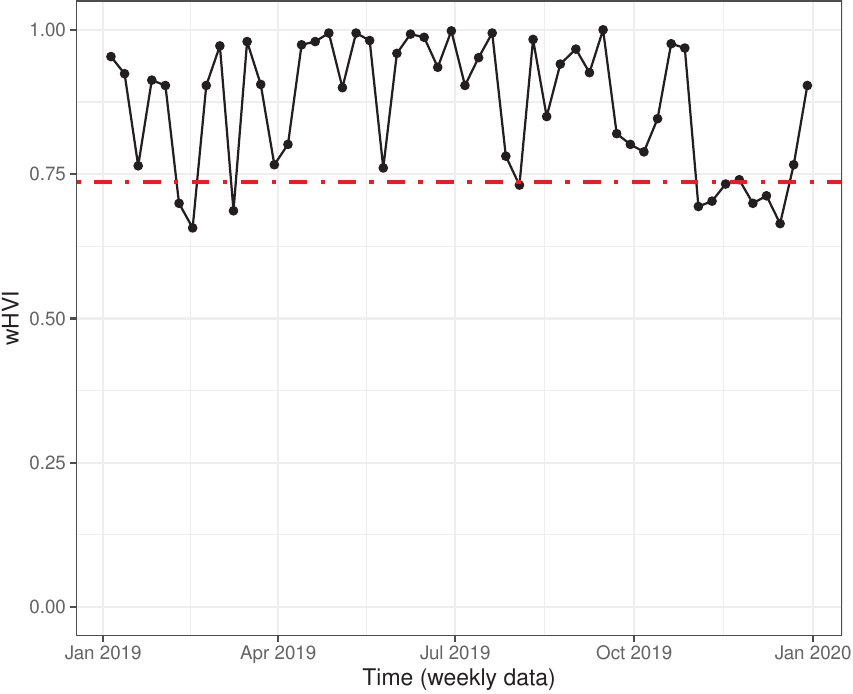}
	\caption{Connected line plot of weekly weighted heat vulnerability index (wHVI) showing fine-scale changes in population vulnerability in 2019 in the Terwent Valley local government area, Tasmania. The annual 2019 vulnerability index is also  displayed as a dashed line to visualise the difference in these relative measures of population vulnerability. Recall that this line is the calculation of the vulnerability index using annual averages and is not equivalent to the average of the monthly index values.}
	\label{fig_vi_2019}
\end{figure}

A range of data collected for the AusEnHealth initiative has enabled the presentation of a finer spatial and temporal scale vulnerability index, allowing for variation in region-specific vulnerability to be observed at the monthly and weekly temporal resolution (see Figures \ref{fig_vi_time} and \ref{fig_vi_2019}). Figure \ref{fig_vi_time} demonstrates how population vulnerability differs when more granular temporal data are considered and compared. For example, the monthly and weekly wHVI values in 2019 both reach the 100$^{th}$ percentile, while the year is represented by a wHVI of $0.74$. In fact, Figure \ref{fig_vi_2019} shows that weekly vulnerability index calculations are often significantly higher than the 2019 annual wHVI for the Terwent Valley LGA. Given the temporal resolution of the underlying data, these intra-seasonal changes are a direct result of fluctuations in exposure, which when combined with the underlying demographics and built environment characteristics over a larger time period, represent reactive population vulnerability at a resolution never seen before across Australia.

It is important to also comment on the utility of the proposed vulnerability index methodology over the general use of all cause mortality for population vulnerability assessments. First, health outcome data can be difficult to attain, especially at fine temporal and geographical resolutions. Health outcome data requests can be overcome by using the more granular underlying characteristics to create fine temporal and geographical vulnerability indices at a national scale, using the proposed correlation-based weights approach to appropriately connect mortality to population vulnerability. Second, the underlying components comprising these vulnerability indices can be explored to identify which components are outliers, thereby contributing significantly to a region's high vulnerability percentile. Finally, the methodology showcased via AusEnHealth is generally applicable globally where relevant spatial data are accessible. Vulnerability indices created using this weighted average approach represent empirical associations with health outcomes used to generate weights and are not necessarily causal. Case-specific epidemiological modelling should be employed to pursue such relationships.

\section{Conclusion}\label{s:Conclusion}
Environmental health vulnerability indices offer a way to assess population vulnerability despite variations in population density and environmental burden of disease. However, common index creation methodology may not accurately reflect timely population health risks. The pairwise correlation approach to vulnerability index methodology allows for a health outcome of interest to guide population vulnerability calculations, while maintaining the critical exposure, sensitivity and adaptive capacity framework used in recent vulnerability index methods. For decision makers, these improved indices offer a data-driven foundation for guiding informed long-term strategic planning, helping to mitigate risks before they escalate into full-blown crises.

The volume of data collected in this study also allows for high temporal and geographical resolutions, allowing users of the data to observe fine changes in environmental exposure and hence determine sharp and sudden changes in population vulnerability due to an emerging natural hazard.

The categories of mortality, as well as the exposure, sensitivity and adaptive capacity variables used in this study reflect the data available at the time of writing. Data included can be greatly expanded given future data availability, including health service availability data, urban environment variables such as road and building density data and upcoming 2021 Australian Census data. There are also opportunities to explore the vulnerability index methodology, including a more rigorous assessment of underlying variables and related health outcomes, as well as addressing variable inter-dependencies.

\section*{Acknowledgements}
We are grateful for the support of the AusEnHealth Steering Committee: Mike Lindsay, Andrea Hinwood, Martine Dennekamp, Phil Delaney, Stuart Barr, Beryl Morris and Claire Sparke. We would also like to acknowledge the guidance and advice provided by AusEnHealth Advisory Board. This group comprises of experts in multidisciplinary fields and are a significant source of knowledge to the project team.

\section*{Funding}
The AusEnHealth initiative has been funded through a collaborative research agreement between FrontierSI, Western Australia Department of Health, Environmental Protection Agency Victoria, Queensland University of Technology (QUT), NGIS, Australian Urban Research Infrastructure Network (AURIN) and the NCRIS Terrestrial Ecosystem Research Network (TERN). We also acknowledge the Healthy Environments and Lives (HEAL) Network, whose ongoing funding has supported the development of this manuscript.

\section*{Abbreviations}
Abbreviations are written as they appear. World Health Organisation (WHO), Australian Environmental Health (AusEnHealth), Bureau of Meteorology (BOM), excess heat factor (EHF), Nitrogen Monoxide (NO), Nitrogen Dioxide (NO2), Sulphur Dioxide (SO2), Carbon Monoxide (CO), Ozone (O3), Particulate Matter $<10\mu m$ (PM10), Particulate Matter $<2.5\mu m$ (PM2.5), Particulate Matter $<1\mu m$ (PM1), Copernicus Atmosphere Monitoring Service (CAMS), Normalised Difference Vegetation Index (NDVI), Australian Bureau of Statistics (ABS), Digital Earth Australia (DEA), Moderate Resolution Imaging Spectroradiometer (MODIS), Terrestrial Ecosystem Network (TERN), Public Health Information Development Unit (PHIDU), mortality over regions and time (MORT), Australian Institute of Health and Welfare, file transfer protocol (FTP), Hypertext Transfer Protocol (HTTP), Australian Statistical Geography Standards (ASGS), statistical area level 2 (SA2), local government area (LGA), Queensland Univerity of Technology (QUT), Intergovernmental Panel on Climate Change (IPCC), social vulnerability index (SVI) social vulnerability index for the United States (SoVI), index of relative socio-economic disadvantage (IRSD), excess cold factor (ECF), principal component analysis (PCA), baseline vulnerability index (VI), weighted vulnerability index (wVI), baseline heat vulnerability index (HVI), weighted heat vulnerability index (wHVI).

\section*{Availability of data and materials}
The data sets associated with this work are provided at an aggregated level and do not contain individual exposures, risk factors or health outcomes. The data are publicly available at \url{https://github.com/QUTCDS/AusEnHealth/Data} and are provided from 2011 through to 2019 at SA2, SA3, SA4 and LGA geographical resolutions and at weekly, monthly and yearly temporal resolutions.

\section*{Authors' contributions}
Data acquisition and processing: Aiden Price and Michael Rigby. Project and stakeholder communication: Paula Fi\'evez and Aiden Price. Technical methodology: Aiden Price and Kerrie Mengersen. Manuscript contributions: Aiden Price, Kerrie Mengersen. All authors contributed to the review of the manuscript.

\bibliographystyle{IEEEtran} 
\bibliography{main}






\section*{Figures and Tables}
\vspace{-1em}

\newpage
\begin{table*}[htbp!]
\centering
\small
	\caption{\textmd{Climate variables collected for use in the AusEnHealth vulnerability indices.}}\label{D_Clim}
	\renewcommand*{\arraystretch}{1.5}
	\begin{tabular}{llll}
		\hline
		\textbf{Variable} \hspace{17em}         & \textbf{Temporal Properties} \hspace{3em} & \textbf{Geographical Properties} \hspace{3em}  & \textbf{Data Source} \\ \hline
		Maximum Temperature ($^{\circ}$C) & 2000-2020 (daily)            & Australia (0.05$^{\circ}$ raster) & BOM                  \\
		Minimum Temperature ($^{\circ}$C) & 2000-2020 (daily)            & Australia (0.05$^{\circ}$ raster) & BOM                  \\ \hline
	\end{tabular}
\end{table*}

\begin{table*}[htbp!]
\centering
\small
	\caption{\textmd{Air quality variables collected for use in the AusEnHealth vulnerability indices.}}\label{D_Air}
	\renewcommand*{\arraystretch}{1.5}
	\begin{tabular}{llll}
		\hline
		\textbf{Variable} \hspace{17em}         & \textbf{Temporal Properties} \hspace{3em} & \textbf{Geographical Properties} \hspace{3em}  & \textbf{Data Source} \\ \hline
		Nitrogen Monoxide, NO (kg kg$^{-1}$) & 2003-2020 (3-hourly)         & Australia (0.75$^{\circ}$ raster) & Copernicus           \\
		Nitrogen Dioxide, NO2 (kg kg$^{-1}$) & 2003-2020 (3-hourly)         & Australia (0.75$^{\circ}$ raster) & Copernicus           \\
		Sulphur Dioxide, SO2 (kg kg$^{-1}$)  & 2003-2020 (3-hourly)         & Australia (0.75$^{\circ}$ raster) & Copernicus           \\
		Carbon Monoxide, CO (kg kg$^{-1}$)   & 2003-2020 (3-hourly)         & Australia (0.75$^{\circ}$ raster) & Copernicus           \\
		Ozone, O3 (kg kg$^{-1}$)             & 2003-2020 (3-hourly)         & Australia (0.75$^{\circ}$ raster) & Copernicus           \\
		Particulate Matter $<10\mu m$, PM10 (kg kg$^{-1}$)                  & 2003-2020 (3-hourly)         & Australia (0.75$^{\circ}$ raster) & Copernicus           \\
		Particulate Matter $<2.5\mu m$, PM2.5 (kg kg$^{-1}$)                 & 2003-2020 (3-hourly)         & Australia (0.75$^{\circ}$ raster) & Copernicus           \\
		Particulate Matter $<1\mu m$, PM1 (kg kg$^{-1}$)                   & 2003-2020 (3-hourly)         & Australia (0.75$^{\circ}$ raster) & Copernicus           \\ \hline
	\end{tabular}
\end{table*}

\begin{table*}[htbp!]
\centering
\small
	\caption{\textmd{Built environment variables collected for use in the AusEnHealth vulnerability indices.}}\label{D_BE}
	\renewcommand*{\arraystretch}{1.5}
	\begin{tabular}{llll}
		\hline
		\textbf{Variable} \hspace{17em}         & \textbf{Temporal Properties} \hspace{3em} & \textbf{Geographical Properties} \hspace{3em}  & \textbf{Data Source} \\ \hline
		Hospital Counts (per population)     & 2021 (yearly)                & Australia (SA2, LGA)              & ABS                  \\
		Green Space (\% parklands)           & 2016 (yearly)                & Australia (SA2, LGA)              & ABS                  \\
		Water Bodies (\%)                    & 2019 (yearly)                & Australia (shapefiles)            & DEA                  \\
		Normalised Difference Vegetation Index              & 2000-2019 (monthly)          & Australia (0.05$^{\circ}$ raster) & MODIS                  \\
		Canopy Cover (\%)                    & 2000-2021 (seasonal)         & Australia (25m grid)              & TERN                  \\ \hline
	\end{tabular}
\end{table*}

\begin{table*}[htbp!]
\centering
\small
	\caption{\textmd{Demographic variables collected for use in the AusEnHealth vulnerability indices.}}\label{D_D}
	\renewcommand*{\arraystretch}{1.5}
	\begin{tabular}{llll}
		\hline
		\textbf{Variable} \hspace{17em}         & \textbf{Temporal Properties} \hspace{3em} & \textbf{Geographical Properties} \hspace{3em}  & \textbf{Data Source} \\ \hline
		Population Density                     & 2014-2019 (yearly)							& Australia (SA2, LGA)              & ABS                  \\
		Median Income                          & 2014-2017 (yearly)							& Australia (SA2, LGA)              & ABS                  \\
		\% Low Income                          & 2016 (yearly)								& Australia (SA2, LGA)              & ABS                  \\
		\% Not Finished High School            & 2011, 2016 (yearly)						& Australia (SA2, LGA)              & ABS                  \\
		Unemployment Rate                      & 2011, 2016 (yearly)						& Australia (SA2, LGA)              & ABS                  \\
		\% Elderly ($>$65 years old            & 2014-2019 (yearly)							& Australia (SA2, LGA)              & ABS                  \\
		\% Infant ($<$5 years old)             & 2014-2019 (yearly)							& Australia (SA2, LGA)              & ABS                  \\
		\% Single Parents                      & 2011, 2016 (yearly)						& Australia (SA2, LGA)              & ABS                  \\
		\% Providing Unpaid Childcare          & 2011, 2016 (yearly)						& Australia (SA2, LGA)              & ABS                  \\
		\% Persons Requiring Core Assistance   & 2011, 2016 (yearly)						& Australia (SA2, LGA)              & ABS                  \\
		\% Persons with a Disability           & 2015, 2018 (yearly)						& Australia (SA2, LGA)              & ABS                  \\
		\% Living Alone                        & 2011, 2016 (yearly)						& Australia (SA2, LGA)              & ABS                  \\
		\% Speaking English as Second Language & 2011, 2016 (yearly)		                & Australia (SA2, LGA)              & ABS                  \\
		\% Indigenous                          & 2011, 2016 (yearly)		                & Australia (SA2, LGA)              & ABS                  \\
		\% Mobile Homes                        & 2014-2019 (yearly)			                & Australia (SA2, LGA)              & ABS                  \\
		\% Homes with Vehicle(s)			   & 2011, 2016 (yearly) 			            & Australia (SA2, LGA)              & ABS                  \\
		\% Crowded Dwellings                   & 2016 (yearly) 					            & Australia (SA2, LGA)              & ABS                  \\
		\% Renters                             & 2011, 2016 (yearly)		                & Australia (SA2, LGA)              & ABS                  \\
		\% Mortgage Payers                     & 2011, 2016 (yearly)		                & Australia (SA2, LGA)              & ABS                  \\
		\% Internet		                       & 2016 (yearly)				                & Australia (SA2, LGA)              & ABS                  \\
		Condition: Respiratory Diseases        & 2011, 2014 (yearly)		                & Australia (LGA)                   & PHIDU                \\
		Condition: Asthma                      & 2011, 2014, 2017 (yearly)	                & Australia (LGA)                   & PHIDU                \\
		Condition: Chronic Obstructive Pulmonary Disease                        & 2011, 2014, 2017 (yearly)	                & Australia (LGA)                   & PHIDU                \\
		Condition: Circulatory System Diseases & 2011, 2014 (yearly)		                & Australia (LGA)                   & PHIDU                \\
		Condition: High Blood Pressure         & 2011, 2014, 2017 (yearly)	                & Australia (LGA)                   & PHIDU                \\
		Condition: High Blood Cholesterol      & 2011, 2014 (yearly)		                & Australia (LGA)                   & PHIDU                \\
		Condition: Cardiovascular Disease      & 2014, 2017 (yearly)		                & Australia (LGA)                   & PHIDU                \\
		Condition: Overweight                  & 2011, 2014, 2017 (yearly)	                & Australia (LGA)                   & PHIDU                \\
		Condition: Obese                       & 2011, 2014, 2017 (yearly)	                & Australia (LGA)                   & PHIDU                \\
		Condition: Smoker                      & 2011, 2014, 2017 (yearly)	                & Australia (LGA)                   & PHIDU                \\
		Condition: High Alcohol Consumption    & 2011, 2014, 2017 (yearly)	                & Australia (LGA)                   & PHIDU                \\ \hline
	\end{tabular}
\end{table*}

\begin{table*}[htbp!]
\centering
\small
	\caption{\textmd{Health outcome variables collected for use in the AusEnHealth vulnerability indices.}}\label{D_HO}
	\renewcommand*{\arraystretch}{1.5}
	\begin{tabular}{llll}
		\hline
		\textbf{Mortality Category} \hspace{12.5em}         & \textbf{Temporal Properties} \hspace{3em} & \textbf{Geographical Properties} \hspace{3em}  & \textbf{Data Source} \\ \hline
		Coronary Heart Disease      & 2014-2019 (5-yearly)         & Australia (SA3)                   & AIHW                 \\
		Cerebrovascular Disease     & 2014-2019 (5-yearly)         & Australia (SA3)                   & AIHW                 \\
		Heart Failure               & 2014-2019 (5-yearly)         & Australia (SA3)                   & AIHW                 \\
		Cardiac Arrhythmia          & 2014-2019 (5-yearly)         & Australia (SA3)                   & AIHW                 \\
		Chronic Obstructive Pulmonary Disease                        & 2014-2019 (5-yearly)         & Australia (SA3)                   & AIHW                 \\
		All Cause                   & 2014-2019 (5-yearly)         & Australia (SA3)                   & AIHW                 \\ \hline
	\end{tabular}
\end{table*}

\begin{table*}[htbp!]
\centering
\small
	\caption{\textmd{Exposure sub-index components used in AusEnHealth to create heat, cold and air quality vulnerability indices.}}\label{T:E}
	\renewcommand*{\arraystretch}{1.5}
\begin{tabular}{llccc}
\hline
\textbf{Theme}                          & \textbf{Variable} & \textbf{Heat} & \textbf{Cold} & \textbf{Air Quality} \\ \hline
\multirow{2}{*}{Heat Exposure}          & Historical Temperature Percentiles (Heat)                      & x             &               &                      \\
                                        & Excess Heat Factor (EHF)                & x             &               &                      \\ \hline
\multirow{2}{*}{Cold Exposure}          & Historical Temperature Percentiles (Cold)                      &               & x             &                      \\
                                        & Excess Cold Factor (ECF)                &               & x             &                      \\ \hline
\multirow{4}{*}{Air Pollutant Exposure} & Nitrogen Monoxide (NO)                  &               &               & x                    \\
                                        & Nitrogen Dioxide (NO2)                  &               &               & x                    \\
                                        & Ozone (O3)                              &               &               & x                    \\
                                        & Particulate Matter (PM2.5)              &               &               & x                    \\ \hline
\end{tabular}
\end{table*}

\begin{table*}[htbp!]
\centering
\small
	\caption{\textmd{Sensitivity sub-index components used in AusEnHealth to create heat, cold and air quality vulnerability indices.}}\label{T:S}
	\renewcommand*{\arraystretch}{1.5}
	\begin{tabular}{llccc}
		\hline
		\textbf{Theme}                         & \textbf{Variable}                                & \textbf{Heat} & \textbf{Cold} & \textbf{Air Quality} \\ \hline
		\multirow{5}{*}{Socio-economic Status} & Population Density (PD)                               & x             & x             & x                    \\
		& Median Income (MI)                                    & x             & x             & x                    \\
		& Low Income (LI)                                       & x             & x             & x                    \\
		& Education (Not Completed High School) (HS)            & x             & x             & x                    \\
		& Unemployment Rate (UR)                               & x             & x             & x                    \\ \hline
		\multirow{7}{*}{Household Composition} & Elderly ($>65$) (ELD)                                 & x             & x             & x                    \\
		& Infants ($<5$) (INF)                                  & x             & x             & x                    \\
		& Single Parents (of Dependants) (SP)                 & x             & x             & x                    \\
		& Unpaid Childcare (UC)                                 & x             & x             & x                    \\
		& Persons Requiring Significant Assistance (ASS)         & x             & x             & x                    \\
		& Persons with a Disability (WD)                       & x             & x             & x                    \\
		& Living Alone (LA)                                     & x             & x             & x                    \\ \hline
		\multirow{2}{*}{Language and Culture}  & English as a Second Language (SL)                    & x             & x             & x                    \\
		& Indigenous (IND)                                       & x             & x             & x                    \\ \hline
		\multirow{4}{*}{Housing Conditions}    & Mobile Homes (MH)                                    & x             & x             & x                    \\
		& Crowded Dwellings (CD)                                & x             & x             & x                    \\
		& Renters (RD)                                         & x             & x             & x                    \\
		& Mortgage Payers (MP)                                 & x             & x             & x                    \\ \hline
		\multirow{7}{*}{Health Status}         & Condition: Respiratory Diseases (RESP)                  &               &               & x                    \\
		& Condition: Asthma (AST)                               &               &               & x                    \\
		& Condition: Chronic Obstructive Pulmonary Disease (COPD) &               &               & x                    \\
		& Condition: Circulatory System Diseases (CSD)          & x             & x             &                      \\
		& Condition: High Blood Pressure (BP)                   & x             & x             &                      \\
		& Condition: High Blood Cholesterol (BC)               & x             & x             &                      \\
		& Condition: Cardiovascular Disease (CD)               & x             & x             &                      \\ \hline
		\multirow{4}{*}{Health Risk Factors}   & Condition: Overweight (OW)                           & x             & x             & x                    \\
		& Condition: Obese (OB)                                & x             & x             & x                    \\
		& Condition: Smoker (SMO)                                & x             & x             & x                    \\
		& Condition: High Alcohol Consumption (ALC)             & x             & x             & x                    \\ \hline
	\end{tabular}
\end{table*}

\begin{table*}[htbp!]
\centering
\small
	\caption{\textmd{Adaptive capacity sub-index components used in AusEnHealth to create heat, cold and air quality vulnerability indices.}}\label{T:A}
	\renewcommand*{\arraystretch}{1.5}
	\begin{tabular}{llccc}
		\hline
		\textbf{Theme}                        & \textbf{Variable}        & \textbf{Heat} & \textbf{Cold} & \textbf{Air Quality} \\ \hline
		Health Services                       & Hospitals (HOSP)                & x             & x             & x                    \\ \hline
		\multirow{3}{*}{Cool Places}          & Greenspace (Parklands) (GS)  & x             & x             & x                    \\
		& Water Bodies (WB)             & x             & x             &                      \\
		& Vegetation Index (NDVI)  & x             & x             & x                    \\ \hline
		\multirow{2}{*}{Social Connectedness} & Dwellings with a Vehicle (NV) & x             & x             & x                    \\
		& Access to Internet (INT)       & x             & x             & x                    \\ \hline           
	\end{tabular}
\end{table*}

\begin{table}[htbp!]
\centering
\small
	\caption{\textmd{Health categories created from the AIHW MORT Books data set.}}\label{T:ACM}
	\renewcommand*{\arraystretch}{1.5}
	\begin{tabular}{ll} \hline
		\textbf{Health Category}\hspace{1em}	& \textbf{Mortality Cause}		\\ \hline
		\multirow{4}{*}{Heat}					& Coronary Heart Disease        \\
		& Cerebrovascular Disease       \\
		& Heart Failure                 \\
		& Cardiac Arrhythmia            \\ \hline
		Air Quality								& Chronic Obstructive Pulmonary \\ \hline
		All Cause								& All Cause						\\ \hline
	\end{tabular}
\end{table}

\end{document}